\input{aipcheck}

\documentclass[
  ,draft            
  ,numberedheadings 
  ]
  {aipproc}

\layoutstyle{6x9}

\begin{document}

\title{ENERGY EXTRACTION FROM BLACK HOLES}


\author{Norbert Straumann}{
  address={Institute for Theoretical Physics, University of Zurich,\\
Winterthurerstrasse 190, CH--8057 Zurich, Switzerland} }

\begin{abstract}
In this lecture I give an introduction to the rotational energy
extraction of black holes by the electromagnetic Blandford-Znajek
process and the generation of relativistic jets. After some basic
material on the electrodynamics of black hole magnetospheres, we
derive the most important results of Blandford and Znajek by making
use of Kerr-Schild coordinates, which are regular on the horizon. In
a final part we briefly describe results of recent numerical
simulations of accretion flows on rotating black holes, the
resulting large-scale outflows, and the formation of collimated
relativistic jets with high Lorentz factors.

\end{abstract}

\maketitle


\section{Introduction}\label{Sect:intro}

For good reasons it is by now generally believed that active
galactic nuclei (AGNs) are powered by supermassive black holes
(BHs), with typical masses of about $10^9~M_{\odot}$. Indeed, this
seems to be the only way to generate the enormous energy of order
$10^{46}~erg/s$ in regions not much larger than the solar system.
Accretion of matter by black holes, offers the most efficient power
supply. A few solar masses of gas per year suffice to power the most
energetic quasars.

It is known for long that AGNs often expel enormous energies in two
oppositely directed relativistic jets with Lorentz factors $\gamma~
\sim 10$. More recently, moderately large Lorentz factors ($\gamma$
a few) have been seen in BH x-ray binaries (XRBs), called \emph{microquasars}.
Relativistic jets have also been observed in gamma-ray bursts
(GRBs). How are these remarkable jets formed? Which mechanisms are
responsible for the concentrated energy input at their origin? Most
probably processes connected to BHs are involved. Among these,
perhaps the most promising possibility is the energy extraction from
a black hole via magnetic fields, since fairly strong magnetic
fields are likely to be present in accretion flows on the central
BH. This scenario, which gives rise to a long-range coherence, is the
main subject of this talk. Basic questions to be
answered are: Which forces drive the jets? What are the mechanisms
that regulate their content, and how are they collimated?

Penrose first discovered that a substantial fraction of a Kerr black
hole mass can be converted, at least in principle, into the energy
of surrounding matter or radiation. However, the Penrose process is
inefficient under typical astrophysical conditions. Later it was
found that external electromagnetic fields can be used to extract
rotational energy of black holes. Of great influence was a
pioneering study by Blandford \& Znajek \cite{9} which demonstrated
the possibility of an electromagnetically driven wind from a
rotating black hole, provided the space around the black hole is
filled with plasma. In their paper they developed a general theory
of force-free steady-state axisymmetric magnetospheres of black
holes, and estimated that the power of the wind could be high enough
to explain the energetics of radio galaxies and quasars.

This work triggered the so called \emph{membrane model} by Thorne
and collaborators \cite{1} (see also \cite{8}). In this model one
first splits the elegant 4-dimensional physical laws of general
relativity (GR) into space and time (3+1 splitting). For a general
situation this can be done in many ways (reflecting the gauge
freedom in GR) since there is no canonical fibration of spacetime by
level surfaces of constant time. (For a stationary BH we shall
choose foliations which are adapted to the corresponding Killing
field.) Relative to these the dynamical variables (electromagnetic
fields, etc) become quantities on an \emph{absolute space} which
evolve as functions of an \emph{absolute time}, as we are accustomed
to from non-relativistic physics. We shall see, for example, that
the 3+1 splitting brings Maxwell's equations into a form which
resembles the familiar form of Maxwell's equations for moving
conductors. We can then use the pictures and our experience from
ordinary electrodynamics.

In a second more specific step one replaces the boundary conditions at the horizon
by physical properties (electric conductivity, etc) of a
\emph{fictitious membrane}. This procedure is completely adequate as
long as one is not interested in fine details \emph{very close} to
the horizon. The details of this boundary layer are, however,
completely irrelevant for astrophysical applications. (The situation
is similar to many problems in electrodynamics, where one replaces
the real surface properties of a conductor and other media by
idealized boundary conditions.)

This approach turns the drawback of the Boyer-Lindquist coordinates,
which become singular at the horizon, into an advantage. The
membrane model has the merit that it allows us to understand
astrophysical processes near black holes more easily, because things
become then closer to the intuition we have gained from other fields
of physics, for instance from the electrodynamics of moving bodies.
However, this membrane reformulation of regularity requirements at the horizon --
in terms of the singular BL coordinates -- is clearly artificial. More
importantly, it hides the fact that the key role in the electrodynamic
Blandford-Znajek mechanism is played not by the black hole event horizon,
but by its ergo-sphere (as in the Penrose process). Later I shall present
an alternative derivation of the most important results of Blandford
and Znajek, in making use of Kerr-Schild coordinates. This foliation is
not singular at the horizon, whence no boundary problems arise.

Kerr-Schild coordinates not only simplify theoretical analysis, but
have also successfully been used in some numerical simulations. In
recent years several codes have been developed to study the magnetic
energy extraction from BHs. In particular, various groups have
addressed the question how relativistic jets may have formed. In the
last part of the lecture I will briefly report on what has been done
and achieved so far in numerical studies of what is now often dubbed
the (generic) ``Blandford-Znajek mechanism'', although only a
specific version was originally proposed.


We begin with some basic material on black hole electrodynamics.

\section{Space-Time Splitting of Electrodynamics}\label{split}

Let us perform the 3+1 splitting of the general relativistic
Maxwell equations on a stationary spacetime
$(M,^{(4)}\!\mathbf{g})$. Most of what follows could easily be
generalized to spacetimes which admit a foliation by spacelike
hypersurfaces (see, e.g., Ref. \cite{2}), but this is not needed in
what follows. Similar 3+1 decompositions can be carried out for the
other equations of general relativistic magnetohydrodynamics (GRMHD) .

Slightly more specifically, we shall assume that globally $M$ is a
product $\mathbf{R} \times \Sigma$, such that the natural coordinate
$t$ of $\mathbf{R}$ is adapted to the Killing field $k$, i.e.,
$k=\partial_{t}$. We decompose the Killing field into normal and
parallel components relative to the ``absolute space''
$(\Sigma,\mathbf{g})$, $\mathbf{g}$ being the induced metric on
$\Sigma$,
\begin{equation}
        \partial_{t} = \alpha\,u + \beta.
        \label{strau:1}
\end{equation}
Here $u$ is the unit normal field and $\beta$ is tangent to
$\Sigma$. This is what one calls the decomposition into lapse and
shift; $\alpha$ is the \emph{lapse function} and $\beta$ the
\emph{shift vector field}. We shall usually work with adapted
coordinates $(x^{\mu})=(t,x^{i})$, where $\{x^{i}\}$ is a coordinate
system on $\Sigma$. Let $\beta = \beta^{i}\partial_{i} \quad
(\partial_{i}=\partial/\partial x^{i})$, and consider the basis of
1-forms
\begin{equation}
        \alpha\,dt, \qquad dx^{i} + \beta^{i}dt.
        \label{strau:2}
\end{equation}
One verifies immediately, that this is dual to the basis
$\{u,\partial_{i}\}$ of vector fields. Since $u$ is perpendicular to
the tangent vectors $\partial_{i}$ of $\Sigma$, the 4-metric has the
form
\begin{equation}
        ^{(4)}\mathbf{g} = -\alpha^{2}dt^{2} +
        g_{ij}\left(dx^{i}+\beta^{i}dt\right)
        \left(dx^{j}+\beta^{j}dt\right),
        \label{strau:3}
\end{equation}
where $g_{ij}dx^{i}dx^{j}$ is the induced metric $\mathbf{g}$ on
$\Sigma$. Clearly, $\alpha$, $\beta$, and $\mathbf{g}$ are all
time-independent quantities on $\Sigma$.

We introduce several kinds of electric and magnetic fields.
Obviously, the spatial 1-forms
\begin{equation}
\check{\mathcal{E}}:=-i_k F, ~~~\check{\mathcal{H}}:= i_k*F,
\label{mod:1}
\end{equation}
have an intrinsic meaning. If we set
\begin{equation}
      F=\mathcal{B}+\check{\mathcal{E}}\wedge dt,~~~*F=\mathcal{D} -
      \check{\mathcal{H}}\wedge dt,
      \label{mod:2}
\end{equation}
the electric and magnetic 2-forms $\mathcal{D},~\mathcal{B}$ are
spatial. Two further electric and magnetic 1-forms are defined by
\begin{equation}
\mathcal{E}=\star\mathcal{D},~~~\mathcal{H}=\star\mathcal{B},
\label{mod:3}
\end{equation}
where $\star$ denotes the spatial Hodge dual on
$(\Sigma,\mathbf{g})$. The corresponding vector fields are denoted
by $\vec{E}$ and $\vec{B}$. These are measured by observers moving
with 4-velocity $u$, so-called FIDOs, for \emph{fiducial observers}.
One easily finds the following algebraic relations:
\begin{equation}
      \check{\mathcal{E}}=
      \alpha\,\mathcal{E}-i_{\beta}\,\mathcal{B},~~~
      \check{\mathcal{H}}=\alpha \mathcal{H} +
      i_{\beta}\,\mathcal{D}.
      \label{mod:4}
\end{equation}

We give also the coordinate components of the various fields:
\begin{equation}
\check{\mathcal{E}}_i=-F_{ti},~~
\check{\mathcal{H}}_i=*F_{ti}=\frac{\alpha}{2}\eta_{ijk}F^{jk},~~
\mathcal{B}_{ij}=F_{ij},~~ B^i=\frac{1}{2}\eta^{ijk}F_{jk},~~
\mathcal{D}_{ij}=* F_{ij}, ~~E^i=\alpha F^{ti} \label{mod:5}
\end{equation}
where $\eta_{ijk}$ denotes the Levi-Civita tensor on
$(\Sigma,\mathbf{g})$.

In terms of the current 4-vector $J^\mu$ the electric charge density
is $\rho_{el}=\alpha J^t$. Beside the spatial current 1-form $j=J_k
dx^k$ we use also its spatial Hodge dual $\mathcal{J}=\star j$.
Using also $\hat{\rho}=\star\rho_{el}$, the Maxwell equations can be
written in 3+1 form as ($\mathbf{d}$ denotes the exterior differential on $\Sigma$):
\begin{eqnarray}
\mathbf{d}\mathcal{B}=0, &  & \partial_{t}\mathcal{B}+\mathbf{d}\check{\mathcal{E}}=0, \nonumber \\
        \mathbf{d}\mathcal{D} = 4\pi \hat{\rho}, &  & -\partial_{t}\mathcal{D}+ \mathbf{d}
        \check{\mathcal{H}} =  4 \pi\check{\mathcal{J}},~~\check{\mathcal{J}}:=\alpha \, \mathcal{J}-i_{\beta}\,\hat{\rho}.
                       \label{mod:6}
\end{eqnarray}
These can be translate into a vector analytic form. (For this as
well as detailed derivations, see \cite{8}.) All this looks like
Maxwell's equations for moving conductors.

\begin{center}
        \textbf{Integral Formulas}
\end{center}

As is well-known from ordinary electrodynamics, it is often useful
to write the basic laws in integral forms. Consider, for instance,
the induction law in (\ref{mod:6}). If we integrate this over a
surface area $\mathcal{A}$, which is \emph{at rest} relative to the
absolute space, we obtain with Stokes' theorem $(\mathcal{C}:=
\partial\mathcal{A})$, using also (\ref{mod:4}),
\begin{equation}
        \oint_{\mathcal{C}}^{}\alpha\,\mathcal{E} =
        -\frac{d}{dt}\int_{\mathcal{A}}^{}\mathcal{B} +
        \oint_{\mathcal{C}}^{}i_{\beta}\,\mathcal{B}.
        \label{mod:7}
\end{equation}
The left hand side is the electromotive force (EMF) along
$\mathcal{C}$. The last term is similar to the additional term one
encounters in Faraday's induction law for moving conductors. It is
an expression of the coupling of $\mathcal{B}$ to the
gravitomagnetic field and plays a crucial role in much that follows.
This term contributes also for a stationary situation, for which
(\ref{mod:7}) reduces to
\begin{equation}
        \textrm{EMF}(\mathcal{C}) =\oint_{\mathcal{C}}^{}\alpha\,\mathcal{E} =
         \oint_{\mathcal{C}}i_{\beta}\,\mathcal{B}.
        \label{strau:26}
\end{equation}

\section{Black Hole in a Homogeneous Magnetic Field}\label{hommag}

As an instructive example and a useful tool we discuss now an exact
solution of Maxwell's equations in the Kerr metric, which becomes
asymptotically a homogeneous magnetic field. This solution can be
found in a strikingly simple manner \cite{3}.

For any Killing field $K$ and its 1-form $K^{\flat}$ one has the following identity $\delta
\,d\,K^{\flat}=2\,R(K)$, where $\delta$ denotes the co-differential
and $R(K)$ is the 1-form with components $R_{\mu \nu}K^{\nu}$. In
components this is equivalent to
\begin{equation}
        K_{\mu \:\:;\alpha}^{\:;\alpha} = -R_{\mu \alpha}\,K^{\alpha}.
        \label{strau:41}
\end{equation}
This form can be obtained by contracting the indices $\sigma$ and
$\rho$ in the following general equation for a vector field
$\xi_{\sigma;\rho \mu} - \xi_{\sigma;\mu \rho} =
\xi_{\lambda}R^{\lambda}_{\sigma \rho \mu}$ and by using the
consequence $K^{\sigma}_{;\sigma}=0$ of the Killing equation
$K_{\sigma;\rho}+K_{\rho;\sigma}=0$. For a vacuum spacetime we thus
have $\delta \,d\,K^{\flat}=0$ for any Killing field. Hence, the
vacuum Maxwell equations are satisfied if $F$ is a constant linear
combination of the differential of Killing fields (their duals, to
be precise). For the Kerr metric, as for any axially symmetric
stationary spacetime, we have two Killing fields $k$ and $m$. Both
in Boyer-Lindquist (BL) and in Kerr-Schild (KS) foliations the
Killing fields are coordinate derivatives: $k=\partial_{t}$ and
$m=\partial_{\varphi}$ ($k^{\flat}$ and $m^{\flat}$ denote the
corresponding 1-forms). The Komar formulae provide convenient
expressions for the total mass $M$ and the total angular momentum
$J$ of the Kerr BH:
\begin{equation}
        M=-\frac{1}{8\pi}\int_{\infty}*dk^{\flat},\qquad J=\frac{1}{16\pi}
        \int_{\infty}*dm^{\flat}
        \label{strau:43}
\end{equation}
(for $G=1$).

We try the ansatz
\begin{equation}
        F = \frac{1}{2}\,B_{0}\,(dm^{\flat}+2a\,dk^{\flat}) \qquad (B_{0} =
        \textrm{const}),
        \label{strau:44}
\end{equation}
and choose $a$ such that the total electric charge
\begin{equation}
        Q = -\frac{1}{4\pi}\int_{\infty}*F
        \label{strau:45}
\end{equation}
vanishes. The Komar formulae (\ref{strau:43}) tell us that
\begin{equation}
        Q = -\frac{1}{8\pi}\,B_{0}\,(16\pi J-2a\cdot 8\pi M),
        \label{strau:46}
\end{equation}
and this vanishes if $a=J/M$ (which is the standard meaning of the
symbol $a$ in the Kerr solution).

Clearly, $F$ is stationary and axisymmetric:
\begin{equation}
        L_{k}\,F = L_{m}\,F = 0,
        \label{strau:47}
\end{equation}
because (dropping $\flat$ from now on) $L_{k}\,dk =d\,L_{k}\,k = 0,~
(L_{k}\,k = [k,k] = 0),\:\textrm{etc}.$

The solution (\ref{strau:44}) can be expressed in terms of a
potential: $F=dA$, with
\begin{equation}
        A = \frac{1}{2}\,B_{0}\,(m+2ak)=\frac{1}{2}\,B_{0}\,(g_{\mu\varphi}+2a\,g_{\mu
        t})\,dx^{\mu},
        \label{strau:55}
\end{equation}
where the last expression holds in Boyer-Lindquist (BL) as well as
in Kerr-Schild (KS) coordinates. Asymptotically, this describes a
magnetic field in the z-direction whose magnitude is $B_{0}$.

It is straightforward to work out explicit expressions for $\vec{E}$
and $\vec{B}$. The electric field has a quadrupole-like structure
and is poloidal. It is proportional to $a$, and thus due to the
gravitomagnetic component of the Kerr solution. Its emergence is of
great astrophysical significance.

\begin{figure}[htbp]
        \centerline{\includegraphics[width=0.4\linewidth]{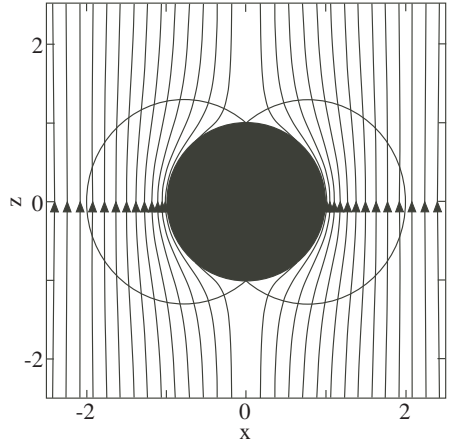}}
        \caption{Black hole ``Meissner effect'' (from \cite{KK}). The magnetic field lines
        of Wald's vacuum solution in KS coordinates for $a=1$ are shown in the
        $x=r\sin\vartheta,~ z=r\cos\vartheta$ plane. The partial circles (thick lines) bound the ergosphere.}
        \protect\label{strau:Fig.3.2}
\end{figure}

It is of interest to work out the magnetic flux through the equator
of the BH, i.e.,
\begin{equation}
        \Phi = \int_{\textrm{upper h.}}\mathcal{B} =
        \int_{\textrm{equator}}\mathcal{A} = 2\pi\,\mathcal{A}_{\varphi}
        \left|_{\textrm{equator}}\right..
        \label{strau:66}
\end{equation}
One finds
\begin{equation}
        \Phi = 4\pi B_{0}\,M(r_{H}-M) =4\pi B_{0}\,M\sqrt{M^{2}-a^{2}}.
        \label{strau:67}
\end{equation}
Generically one has, as expected, $\Phi \approx \pi r_{H}^{2}B_{0}.$
Note, however, that (\ref{strau:67}) vanishes for an extremal BH
($a=M$). In other words, the flux is completely expelled from the
black hole, like in the Meissner-Ochsenfeld effect in
superconductivity. In order to study the structure the Wald solution
close to the black hole one has to express it in terms of
Kerr-Schild coordinates (see Sect. \ref{b-z}). The magnetic field
lines in these coordinates are shown for $a=M$ in Fig.
\ref{strau:Fig.3.2}, taken from \cite{KK}.

 \section{Axisymmetric stationary Fields}

In this section we discuss some  consequences of Maxwell's equations
for a stationary axisymmetric magnetoshere outside a black hole.

\subsection{Potential Representation}

\begin{figure}[htbp]
        \centerline{\includegraphics[width=0.4\linewidth]{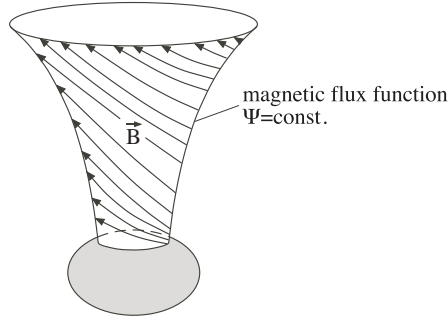}}
        \caption{Axisymmetric magnetic field. The total flux inside the
        magnetic surface defines the flux function $\Psi$.}
        \protect\label{strau:Fig.5.2}
\end{figure}

For an axisymmetric field we have
$L_{\partial_{\varphi}}\,\mathcal{B}=0\leftrightarrow
\mathbf{d}i_{\partial_{\varphi}}\,\mathcal{B}=0$, whence
$i_{\partial_{\varphi}}\,\mathcal{B}= -\mathbf{d}\Psi/2\pi$. From
this we conclude that for the BL foliation $\mathcal{B}$ can be
expressed in terms of two potentials $\Psi$ and $I$,
\begin{equation}
        \mathcal{B} =\underbrace{ \frac{1}{2\pi}\, \mathbf{d}\Psi \wedge \mathbf{d}\varphi}_{\textrm
        {poloidal part}} + \underbrace{\frac{2I}{\alpha}\mathbf{*}\mathbf{d}\varphi}_{\textrm{toroidal part}},
        \label{strau:5.7}
\end{equation}
both of which can be taken to be independent of $\varphi$. (The
physical meaning if $I$ will be discussed further below.) $\Psi$ is
the magnetic flux function (see Fig. \ref{strau:Fig.5.2}), because
the poloidal flux inside a tube $\{\Psi = \textrm{const}\}$ is
\begin{equation}
        \int \mathcal{B} = \frac{1}{2\pi}\int \mathbf{d}(\Psi
        \mathbf{d}\varphi)=\frac{1}{2\pi}\oint\Psi \mathbf{d}\varphi = \Psi, \qquad \Psi(0)=0.
        \label{strau:5.9}
\end{equation}
$\Psi$ is constant along magnetic field lines, as should be clear
from Fig. \ref{strau:Fig.5.2}. It is easy to show this also
formally.

The electric field $\vec{E}$ has no toroidal component for an
axisymmetric stationary situation. This is an immediate consequence
of the induction law: Applying (\ref{mod:7}) for a stationary and
axisymmetric configuration to the closed integral curve
$\mathcal{C}$ of the Killing field $\partial_\varphi$, we obtain
\begin{equation}
    \oint_{\mathcal{C}}\alpha\,\mathcal{E}
    =\oint_{\mathcal{C}}i_{\beta}\,\mathcal{B}
    =\frac{\omega}{2\,\pi}\oint_{\mathcal{C}}\mathbf{d}\Psi =
    0 \Longrightarrow \vec{E}^{\textrm{tor}}=0.
        \label{strau:8.2}
\end{equation}
A similar application of Amp\`{e}re's law in integral form gives
\begin{equation}
        \oint_{\mathcal{C}}\alpha\mathcal{H} = 4\,\pi
        \int_{\mathcal{A}}\alpha\mathcal{J} = 4\,\pi\,I,
        \label{strau:8.3}
\end{equation}
where $I$ is the total upward current through a surface
$\mathcal{A}$ bounded by $\mathcal{C}$. This shows that the
potential $I$ in (\ref{strau:5.7}) is the upward current.

So far we did not make any model assumptions about the physics of
the magnetosphere. But now we assume that the electromagnetic field
is \emph{degenerate}, i.e., that $\vec{E}\cdot\vec{B}=0$, which is
equivalent to the invariant statement $F\wedge F=0$. (This is
satisfied for ideal GRMHD and for force-free fields; see Sects.
\ref{grmhd} and \ref{b-z}.) Because $\mathcal{E}$ is poloidal, we
can then represent the electric field as follows
\begin{equation}
        \mathcal{E} = i_{\vec{v}_{F}}\,\mathcal{B} \qquad (\vec{E} =
        -\vec{v}_{F} \times \vec{B}),
        \label{strau:8.9}
\end{equation}
where $\vec{v}_{F}$ is toroidal. Let us set
\begin{equation}
        \vec{v}_{F} =: \frac{1}{\alpha}\,(\Omega_{F}-\omega)\,\tilde{\omega}\,\vec{e}
        _{\varphi}.
        \label{strau:8.10}
\end{equation}
For the interpretation of $\Omega_{F}$ note the following: For an
observer, rotating with angular velocity $\Omega$, the 4-velocity is
$u=u^{t}\,(\partial_{t}+\Omega\,\partial_{\varphi})$. On the other
hand, $u=\gamma\,(e_{0}+\vec{v})$, where $\vec{v}$ is the 3-velocity
relative to a FIDO. Using also
$\partial_{t}=\alpha\,e_{0}+\vec{\beta}$ we get
$\Omega\,\vec{\partial}_{\varphi}=\alpha\, \vec{v}-\vec{\beta}$ or
\begin{equation}
        \vec{v} =
        \frac{1}{\alpha}\,(\Omega-\omega)\,\vec{\partial}_{\varphi} =
        \frac{1}{\alpha}\,(\Omega-\omega)\,\tilde{\omega}\,\vec{e}_{\varphi}.
        \label{strau:8.11}
\end{equation}
This has the same form as (\ref{strau:8.10}). Since the transformed
electric field $\vec{E}'=\gamma_F(\vec{E}+ \vec{v}_{F} \times
\vec{B})$ vanishes, we may regard $\Omega_{F}$ as the angular
velocity of the magnetic field lines. (The transformed frame can be
defined as the ``local rest frame'' of the magnetic field lines.)
Eqs. (\ref{strau:8.9}) and (\ref{strau:8.10}) imply
\begin{equation}
        \alpha \, \mathcal{E} =
        -\frac{\Omega_{F}-\omega}{2\,\pi}\,\mathbf{d}\Psi,~~~\check{\mathcal{E}}=-\mathbf{d}\Psi,
        \label{strau:8.12}
\end{equation}
thus $\vec{E}$ is \emph{perpendicular} to the surfaces $\{\Psi =
\textrm{const}\}$. Taking the exterior derivative, and using
induction law, we get $\mathbf{d}\Omega_{F} \wedge \mathbf{d}\Psi =0
\Longrightarrow \Omega_{F} = \Omega_{F}(\Psi)$. So the
electromagnetic field is determined in terms of $\Psi,~I$ and the
``flow (field line) constant'' $\Omega_F(\Psi)$.

\subsection{EMF outside a rotating Black Hole}\label{emf}

\begin{figure}[htbp]
        \centerline{\includegraphics[width=0.4\linewidth]{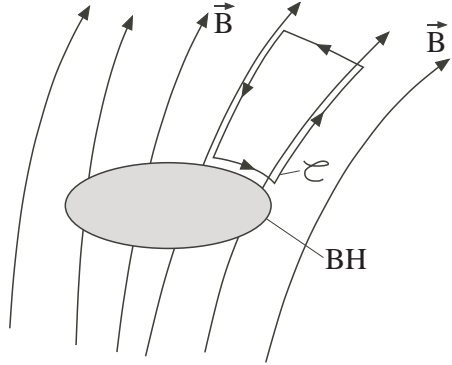}}
        \caption{Arrangement for eq. (\ref{strau:5.3}).}
        \protect\label{strau:Fig.5.1}
\end{figure}

In Fig. \ref{strau:Fig.5.1} we consider a stationary rotating BH in
an external magnetic field (like in \S 3). The integral in
(\ref{strau:26}) along the field lines gives no contribution and far
away $\beta$ drops rapidly ($\sim r^{-2}$). Thus, there remains only
the contribution from the horizon ($\mathcal{C}_{H}$) of the path
$\mathcal{C}$ in Fig. \ref{strau:Fig.5.1}:
\begin{equation}
        \textrm{EMF} = \int_{\mathcal{C}_{H}}^{}i_{\beta_{H}}\,\mathcal{B}, \quad
        \beta_{H}=-\Omega_{H}\,\partial_{\varphi},
        \label{strau:5.3}
\end{equation}
where $\Omega_H$is the angular velocity of the horizon (only the
normal component $\vec{B}_{\perp}$ contributes). I recall that
$\Omega_{H}=a(2Mr_{H})^{-1},\ r_{H}=M+\sqrt{M^{2}-a^{2}}$.

Let us work this out for the special case of an axisymmetric field.
For the closed path $\mathcal{C}$ in Fig. \ref{strau:Fig.5.1} the
EMF is by (\ref{strau:5.3})
\begin{displaymath}
        \textrm{EMF} \equiv \triangle V = \int_{\mathcal{C}_{H}}^{}i_{\beta_{H}}\,\mathcal{B}
        = -\left(-\frac{\Omega_{H}}{2\pi} \right)\,
        \int_{\mathcal{C}_{H}}\mathbf{d}\Psi,
\end{displaymath}
i.e.
\begin{equation}
        \triangle V = \frac{\Omega_{H}}{2\pi}\,\triangle \Psi.
        \label{strau:5.10}
\end{equation}
This result is \emph{independent} of the physics outside the black
hole.

Let us integrate it from the pole to some point north of the
equator. For the exact vacuum solution in \S 3 we know the result
for the EMF, if we integrate up to the equator: From
(\ref{strau:5.10}) and (\ref{strau:67}) we get $\textrm{EMF} =
2\Omega_{H}\,B_{0}\,M\,(r_{H}-M)$ or $(\Omega_{H}=a/2Mr_{H})$
\begin{equation}
        \textrm{EMF}  = a\,B_{0}\,\frac{r_{H}-M}{r_{H}}\qquad
        \left(r_{H}=M+\sqrt{M^{2}-a^{2}}\right).
        \label{strau:5.15}
\end{equation}
Note that the ``Meissner effect'' for black holes implies that the
\emph{EMF vanishes for extremal black holes}. We shall, however, see
that this property of the vacuum solution is astrophysically not
relevant, because a plasma-filled magnetosphere changes the
structure of the magnetic field close to the horizon dramatically.
Even for a maximally rotating Kerr black hole the magnetic field is
pulled inside the event horizon. For a detailed discussion we refer
to \cite{KK}.

For a general situation we have roughly $\Sigma\,\triangle \Psi=\Psi
\sim B_{\perp}\,\pi\,r_{H}^{2},~ \tilde{\omega}^{2} \sim \
<\tilde{\omega}^{2}>\  \sim \frac{r_{H}^{2}}{2}$. The total EMF,
$V=\Sigma\,\triangle V$, is thus
\begin{equation}
        V \sim \frac{1}{2\pi}\,\Omega_{H}\,\Psi \sim \frac{1}{2\pi}\,
        \frac{a}{2Mr_{H}}\,B_{\perp}\,\pi\,r_{H}^{2} \simeq \frac{1}{2}
        \left(\frac{a}{M}\right)MB_{\perp}
        \label{strau:5.17}
\end{equation}
(compare this with (\ref{strau:5.15})). Numerically we find
\begin{equation}
        V \sim (10^{20}\,\textrm{Volt}) \left(\frac{a}{M}\right)\,
        \frac{M}{10^{9}\,M_{\odot}}\,\frac{B_{\perp}}{10^{4}\,G}.
        \label{strau:5.18}
\end{equation}
For reasonable astrophysical parameters we obtain magnetospheric
voltages $V$ $ \sim 10^{20}$ Volts. This voltage is comparable to
the highest cosmic ray energies that have been detected.

Note, however, that for a realistic astrophysical situation there is
plasma outside the BH and it is, therefore, at this stage not clear
how the horizon voltage (\ref{strau:5.18}) is used in accelerating
particles to very high energies. This crucial issue is addressed in
the final section.

Let us estimate at this point the characteristic magnetic field
strength than can be expected outside a supermassive BH. A characteristic measure
is the field strength $B_{E}$, for which the energy density $B_{E}^{2}/8\pi$ is
equal to the radiation energy density $u_{E}$ corresponding to the Eddington luminosity. One finds
$(M_{H,8} \equiv M_{H}/10^{8}\,M_{\odot})$
\begin{equation}
        B_{E} = 1.2 \times 10^{5}\,M_{H,8}^{-1/2}\ \textrm{Gauss}.
        \label{strau:5.22}
\end{equation}
For a BH with mass $\sim 10^{9}\ M_{\odot}$ inside an accretion disk
acting as a dynamo, a characteristic field of about 1 Tesla
($10^{4}$ Gauss) is thus quite reasonable.

\section{Basic equations of general relativistic ideal
magnetohydrodynamics}\label{grmhd}

The relativistic fluid is described by its rest-mass density,
$\rho_0$, the energy-mass density, $\rho$, the 4-velocity, $U^\mu$,
and the isotropic pressure, $p$, assumed to be given by an ideal gas
equation of state
\begin{equation}
       p=(\Gamma-1)\varepsilon,
       \label{mhd:6.1}
\end{equation}
where $\varepsilon=\rho-\rho_0$ is the internal energy density, and
$\Gamma$ is the adiabatic index.

The basic equations of GRMHD are easy to write down. First, we have
the baryon conservation
\begin{equation}
       \nabla_\mu(\rho_0 U^\mu)=0.
       \label{mhd:6.2}
\end{equation}
For a magnetized plasma the equations of motion are
\begin{equation}
       \nabla_\nu T^{\mu\nu}=0,
       \label{mhd:6.3}
\end{equation}
where the energy-stress tensor $T^{\mu\nu}$ is the sum of the matter
(M) and the electromagnetic (EM) parts:
\begin{eqnarray}
      T^{\mu\nu}_{M}&=&(\rho+p)U^\mu U^\nu +pg^{\mu\nu}, \label{mhd:6.4}\\
      T^{\mu\nu}_{EM} &=&\frac{1}{4\pi}(F^\mu{}_\lambda
      F^{\nu\lambda}-\frac{1}{4}g^{\mu\nu}
      F_{\alpha\beta}F^{\alpha\beta}).
      \label{mhd:6.5}
\end{eqnarray}
In addition we have Maxwell's equations
\begin{equation}
      dF=0,~~~~ \nabla_\nu F^{\mu\nu}=4\pi J^\mu.
      \label{mhd:6.6}
\end{equation}
We adopt the \emph{ideal MHD approximation}
\begin{equation}
      i_U F=0,
      \label{mhd:6.7}
\end{equation}
which implies that the electric field vanishes in the rest frame of
the fluid (infinite conductivity). Then the inhomogeneous Maxwell
equation provides the current 4-vector $J^\mu$, but is otherwise not
used in what follows.

As a consequence of (\ref{mhd:6.6}) and (\ref{mhd:6.7}) we obtain
(using the Cartan identity $L_U=i_U\circ d + d\circ i_U$): $L_U
F=0$, i.e., that $F$ is invariant under the plasma flow, implying
flux conservation. The basic equations imply that
\begin{equation}
*F= \mbox{\boldmath$h$}\wedge U, \label{MHD:1}
\end{equation}
where $\mbox{\boldmath$h$}=i_U *F$ is the magnetic induction in the
rest frame of the fluid (seen by a comoving observer). Note that
$i_U \mbox{\boldmath$h$}=0$. Furthermore, one can show that the
electromagnetic part of the energy-momentum tensor may be written in
the form
\begin{equation}
T^{\mu\nu}_{EM}=\frac{1}{4\pi}\left[\frac{1}{2}\|\mbox{\boldmath$h$}\|^2g^{\mu\nu}
+\|\mbox{\boldmath$h$}\|^2 U^\mu U^\nu -h^\mu h^\nu \right],
\label{MHD:2}
\end{equation}
with $\|\mbox{\boldmath$h$}\|^2=h_\alpha h^\alpha$.

Let us also work out the 3+1 decomposition of the ideal MHD
condition (\ref{mhd:6.7}). Using $U=\gamma(e_0
+\vec{v}),~\gamma=(1-v^2)^{-1/2}$ and the coordinate velocity
$V^i=U^i/U^t=\alpha v^i-\beta^i$, we obtain from (\ref{mod:2})
\begin{equation}
        \mathcal{E}=i_{\vec{v}}\mathcal{B} \quad\mbox{or}\quad  \check{\mathcal{E}}
        =i_{\vec{V}}\mathcal{B}~~(\Rightarrow i_{\vec{V}}\check{\mathcal{E}}=0).
        \label{mhd:6.7a}
\end{equation}
Therefore, the induction equation becomes
\begin{equation}
\partial_{t}\mathcal{B}+d i_{\vec{V}}\mathcal{B}=0 \quad\mbox{or}\quad \partial_{t}\mathcal{B}+
L_{\vec{V}}\mathcal{B}=0. \label{mhd:6.7b}
\end{equation}

\section{The Blandford-Znajek Process}\label{b-z}

In this section we derive the main results obtained by Blandford and
Znajek \cite{9}, by making use of the Kerr-Schild coordinates.

\subsection{Steady-state force-free magnetospheres}

BZ studied axisymmetric, force-free magnetized plasmas outside Kerr
black holes. The following presentation is strongly influenced by
the recent paper \cite{10}.

A plasma is said to be \emph{force-free} if $i_J F=0$, i.e.
\begin{equation}
        F_{\mu\nu}J^\nu=0
        \label{mhd:6.12a}
\end{equation}
(no electric field in the rest system of the current). This
condition follows from ideal GRMHD when the inertia of the plasma is
ignored. Formally, it is obtained by letting the specific enthalpy
$(\rho+p)/n$ in the ideal GRMHD equations go to zero. Since
Maxwell's equations imply that $\nabla_\nu
T^{\mu\nu}_{EM}=-F^\mu{}_\nu J^\nu$, the energy-stress tensor of the
electromagnetic field is separately conserved. The force-free
condition (\ref{mhd:6.12a}) implies the constraint $F\wedge F=0$,
i.e., the algebraic condition $\ast F_{\mu\nu} F^{\mu\nu}=0.$ In the
literature it has often been asserted that magnetospheres of black
holes should in large parts be nearly force-free This is not born
out in recent simulations, except in the polar region (see Sect.
\ref{sim}). In 3+1 decomposition (\ref{mhd:6.12a}) becomes

\begin{equation}
        \rho_{e}\,\vec{E} + \vec{j} \times \vec{B} = 0.
        \label{strau:8.1}
\end{equation}
Therefore, the vector fields $\vec{E}$ and $\vec{B}$  are
perpendicular.

\subsection{Energy flux at infinity}

Of particular interest is the energy flux at infinity. Since
$\partial_t$ is a Killing field, $T_t^\mu$ are the components of a
conserved 4-vector field:
\[
        \frac{1}{\sqrt{-g}}\partial_\mu(\sqrt{-g}T_t^\mu)=0.
\]
For stationary fields $\partial_i(\sqrt{-g}T_t^i)=0$. The
electromagnetic power at infinity\footnote{In the classical BZ
process one ignores the extraction of rotational energy of the black
hole by accreting material interacting with the electromagnetic
field. Below $T^{\mu\nu}$ always denotes the electromagnetic
energy-momentum tensor.} is
\begin{equation}
        P_{EM}=\int_0^\pi \sqrt{-g}F_E\, 2\pi\,d\vartheta,
        \label{mhd:6.21}
\end{equation}
where $F_E=-T_t^r$. The integral can be taken for any fixed $r$
outside the horizon, in particular at the horizon if we use
Kerr-Schild coordinates. For $F_E$ we get
\begin{equation}
        F_E=-\frac{1}{4\pi}F_{tl}F^{rl}=-\frac{1}{4\pi}F_{t\vartheta}F^{r\vartheta}
        \label{mhd:6.22}
\end{equation}
($F_{t\varphi}=0$). It is straightforward to check that $F_E$ is
invariant under the coordinate transformation [BL]$\rightarrow$
[KS]. Using previous results on readily finds in BL coordinates
\begin{equation}
        F_E=-\frac{1}{4\pi}\Omega_F\frac{(-g)}{g_{rr}g_{\vartheta\vartheta}}\hat{B}^r\hat{B}^\varphi,
        \label{mhd:6.26}
\end{equation}
where $\hat{B}^i:=B^i/\alpha=*F^{it}$.

Now we transform the result (\ref{mhd:6.26}) to KS coordinates. It
is straightforward to show that $\hat{B}^r,~\hat{B}^{\vartheta}$
remain invariant, while the remaining component transforms as
\begin{equation}
        \hat{B}^\varphi[BL]=\hat{B}^\varphi[KS]- \frac{a-2r\Omega_F}{\Delta}
        \hat{B}^r[KS],
        \label{mhd:6.28}
\end{equation}
where $\Delta:=r^2-2Mr+a^2$. This implies that on the horizon ($\Delta=0$)
\begin{equation}
        \left. F_E
        \right|_H=\frac{1}{2\pi}\left(\hat{B}^r\right)^2\Omega_F r_H (\Omega_H-\Omega_F)\sin^2\vartheta.
        \label{mhd:6.30}
\end{equation}
Therefore,
\begin{equation}
        P_{EM}=\int_0^\pi d\vartheta\rho_H^2\sin \vartheta \left(\hat{B}^r\right)^2
        \Omega_Fr_H (\Omega_H-\Omega_F)\sin^2\vartheta
        \label{mhd:6.31}
\end{equation}
or in terms of the normalize normal component $B_\bot$
\begin{equation}
        P_{EM}=\frac{1}{2}\int_0^\pi
        d\vartheta\Omega_F(\Omega_H-\Omega_F)\rho_H\tilde{\omega}^3
        B_\bot^2.
        \label{mhd:6.32}
\end{equation}
The important result (\ref{mhd:6.32}) shows that $P_{EM}$ becomes maximal for
$\Omega_F\approx \frac{1}{2}\Omega_H$. Then
\begin{equation}
        P^{max}_{EM}=\frac{1}{8}\Omega_H^2\int_0^\pi\rho_H\tilde{\omega}^3
        B_\bot^2~d\vartheta.
        \label{mhd:6.34}
\end{equation}
After angular integration one finds, if $B^2_\perp$ is replaced by an
average value $\langle B_\perp\rangle^2$,
\begin{equation}
        P^{max}_{EM}=\frac{1}{4}\left(\frac{a}{M}\right)^2M^2\langle
        B_\perp\rangle^2f\left(\frac{a}{r_H}\right),
        \label{mhd:6.35}
\end{equation}
where the function $f$ is not far from 1 \cite{LWB} . Numerically,
\begin{equation}
        P^{max}_{EM}= 1.7\times 10^{46}~\frac{erg}{s}\left(\frac{a}{M}\cdot\frac{M}{10^9~M_\odot}\cdot
        \frac{\langle B_\perp\rangle}{10^4~G}\right)^2 f.
        \label{mhd:6.37}
\end{equation}

For the angular momentum flux $F_L$ one finds $F_E=\Omega_F F_L$.
Using this one can show that up to 9\% of the initial mass can, in
principle, be extracted \cite{LWB}.

Whether the crucial condition $\Omega_F\approx \frac{1}{2}\Omega_H$
is approximately satisfied in realistic astrophysical scenarios is a
difficult problem for model builders. This brings me to recent
numerical work by several groups.

\section{General Relativistic MHD Simulations}\label{sim}

In recent years several groups have developed codes for ideal GRMHD,
and applied them in particular for studies of the generic BZ
mechanism. These numerical studies show how accretion dynamics
self-consistently create large scale magnetic fields and explore the resulting
outflows.

\subsection{Numerical methods}

There are many ways to write the basic equations in a form suitable
for numerical integration. A pioneering code, based on the 3+1
splitting relative to the FIDO tetrad described in Sect. 2, has been
developed by Koide and collaborators (for a detailed description,
see \cite{Ko}). The system to be integrated consists of eight
evolutionary equations for two scalar and two vector quantities
(primary code variables), which contain in addition five `primitive'
variables that can be determined from the former by solving two
coupled polynomial equations $P(x,y)=0, ~Q(x,y)=0$. This process,
like similar iterative root-finding processes in other schemes, is
time consuming.

A code that is able to perform long-term (several thousand of $M$ in
time) GRMHD simulations, has been developed by Villiers \& Hawley
\cite{VH}. It evolves different auxiliary variables from which the
primitive variables are easily recovered. In contrast to some
alternative schemes \cite{Harm}, \cite{Kom}, it is not fully
conservative. Since BL coordinates are used, the inner boundary
condition must lie outside the horizon. KS coordinates are used in
the axisymmetric code HARM \cite{Harm}. In a modified version of
this \cite{NG} the inversion from ``conserved'' quantities to
``primitive'' variables has been improved (see Appendix A).

A specific common problem is that the magnetic field is advanced in
time by an antisymmetric differential operator, and not by a
differential operator of divergence form. In addition one has to
guarantee that the numerical scheme preserves the constraint
equation $d\mathcal{B}=0$ to rounding error. One way to handle this
problem is the so-called constraint transport (CT) method, where the
induction equation is discretized such that the solenoidal
constraint is built in \cite{EH}.

No code is perfect, and it is therefore important that the outcomes
of different approaches are compared.

\subsection{Qualitative results}

I first summarize some of the results presented in \cite{HK} on
simulations of accretion flows on rotating black holes, and the
properties of the resulting unbound outflows\footnote{This is based
on a sequence of earlier papers on the the subject, cited in
\cite{HK}}.

\begin{figure}[htbp]
        \centerline{\includegraphics[width=0.4 \linewidth]{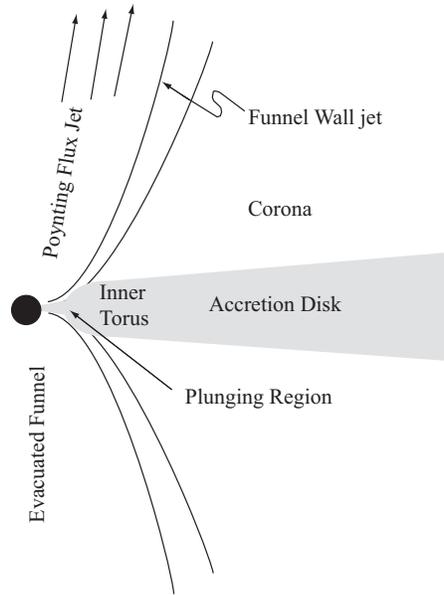}}
        \caption{Main dynamical features in accretion disk simulations.
        Fig. 1 of \cite{HK}.}
        \protect\label{mhd:Fig.8.1}
\end{figure}

The qualitative late-time structure is illustrated in Fig.
\ref{mhd:Fig.8.1}. Along the equator there is a wedge-shaped
Keplerian accretion disk and a net accretion flow is produced by MHD
turbulence whose origin is a magneto-rotational instability, which
leads to a fast amplification of the magnetic field (see Appendix
B). A short distance outside the marginally stable circular stable orbit,
the equatorial pressure and density reach a local maximum in a
`inner torus' region. Inside this local pressure maximum, the
density and pressure drop as the flow accelerates toward the black
hole in the `plunging region'. Above and below the disk is a
`corona' of hot magnetized plasma with a magnetic field whose
typical strength is near equipartition. Along the spin axis of the
BH there is a `centrifugal funnel' that is largely empty of matter,
but filled with magnetic field and an outward Poynting flux. Between
the evacuated funnel and the corona there is a region of unbounded
mass flux, referred to as the `funnel wall jet'. It is this part
that is the center of attention in \cite{HK}. Of particular interest
is the strength of the jet compared to the amount of accretion on
the BH. The quantification of this is somewhat ambiguous, because
matter and electromagnetic fluxes are usually not constant in time
or radius, and there is exchange between the two components.

The numerical results show that for rapidly-rotating BH's the jet
plays a significant role in the energy budget. When $\mid a/M\mid
\,\geq0.9$, the total jet efficiency is generally several tens of
percent, with the matter portion somewhat larger than the
electromagnetic part. The ratio of \emph{angular momentum} in the
unbound outflow to that deposited in the BH is also a very strong
function of BH spin. In the non-rotating case this is only a few
tenths of a percent, but rises to $\simeq 25\%$ when $a/M=0.9$. For
$a/M=0.99$, the highest spin case reported in \cite{HK}, the rate at
which the BH spins down due to electromagnetic torques nearly
matches the rate at which it acquires angular momentum by accreting
matter. Indeed, the angular momentum expelled in the outflow is an
order of magnitude larger than the net amount captured. In all
cases, the electromagnetic portion of the angular momentum carried
away is comparable to the matter portion carried in the funnel wall
jet.

It has to be emphasized that in the initial state of the simulations
there is no magnetic field in the region that eventually becomes the
outflow region. The large scale magnetic field within the funnel and
the Poynting flux jet rapidly develop as a result of magnetically
controlled accretion. The formation can be described as follows.
When disk material reaches the horizon, strong magnetic pressure
gradients are built up which drive the plasma upward. In turn, this
motion drains off the field lines, forming a ``magnetic tower''.
This was previously seen in non-relativistic simulations using a
pseudo-Newtonian potential \cite{KMS}. It is interesting that the
evacuated funnel is the only region that is force-free\footnote{The
criterion for the validity of this property is that
$(\|h\|^2/8\pi)/(\rho+p)\gg 1$.} (in contrast to earlier
expectations).

The matter-dominated outflow moves at a modest velocity ($v/c\sim
0.3$) along the centrifugal barrier surrounding the evacuated
funnel. The funnel wall jet turns out to be \emph{accelerated} and
\emph{collimated} by magnetic and gas pressure forces in the inner
torus and the surrounding corona. The magnetic field is spun by the
rotating Kerr spacetime, hence the energy of the Poynting flux jet
comes from the BH rotation. Below we will address the question
whether magnetic forces might provide additional acceleration and
collimation on far larger scales than are modeled in \cite{HK}.

Hawley and Krolik conclude from their numerical studies that while
the proximate energy source is the BH's rotation, accretion
\emph{replenishes} both the BH's mass and angular momentum. A
substantial decrease of the rotational energy of the BH does not
appear to be a generic phenomenon. This is in contrast to the
classic BZ model. For a detailed discussion, I refer to the original
paper.

Similar results were found before by McKinney and Gammie in their
axisymmetric simulations \cite{10}. Among other aspects these also
showed that as the hole loses energy and angular momentum, its total
mass and angular momentum are replenished by accretion. In
\cite{McK} McKinney followed the evolution of the jet till $t\approx
10^4 GM/c^3$ out to $r\approx 10^4GM/c^2$, and found that by then
the jet has become superfast magnetosonic and moves at a Lorentz
factor of about 10. This may, however, only be a small fraction of
the Lorentz factor at much larger distances. Indeed, if a large
fraction of the magnetic and thermal energy would go into particle
kinetic energy, one would estimate from the simulation that the
terminal Lorentz factor may reach almost $10^3$. This estimate is
based on the following fact. For a stationary axisymmetric flow the
energy (momentum) flux per unit rest-mass flux is conserved along
flux surfaces and can thus be determined from local flow quantities.
The relevant formula for this quantity is, using previous notation,
\begin{equation}
    \frac{-T^A_t}{\rho_0 U^A}=-\frac{\rho+p}{\rho_0}U_t
    -\frac{\Omega_F}{4\pi}\frac{\hat{B}^A}{\rho_0
    U^A}\check{\mathcal{H}}_\varphi ~~~(A=r,\vartheta). \label{sim:1}
\end{equation}

Since the simulation in \cite{McK} stopped long before the end of
the acceleration period, Komissarov et al. have studied numerically
the further evolution of the relativistic jet \cite{Kom2}. Key
issues of this investigation are: (i) Is the magnetic driving
mechanism able to accelerate outflows to high Lorentz factors with
high efficiency over astrophysically extended scales? (ii) Can these
flows be collimated by purely magnetic stresses? Since most of the
acceleration takes place far away from the black hole, the
simulations are carried out in the framework of special-relativistic
ideal MHD.

The authors investigate models of the following kind. The free
boundary with an ambient medium is replaced by solid rigid walls on
which appropriate boundary conditions are imposed. This
simplification enables higher numerical accuracy. At the inlet
boundary, the injected poloidal current distribution is prescribed
through a rotational profile. The initial configuration corresponds
to a non-rotating purely poloidal magnetic field with nearly
constant magnetic pressure across the funnel. Moreover, the outflows
are initially Poynting flux-dominated. The authors find that these
approach a steady state with a spatially extended acceleration
region. Furthermore, the acceleration process turns out to be very
efficient; almost 80\% of the Poynting flux is converted into
kinetic energy. The results also show efficient self-collimation. In
contrast to \cite{McK}, no instabilities or shocks are found in the
simulation.

\begin{figure}[htbp]
        \centerline{\includegraphics[width=0.6\linewidth]{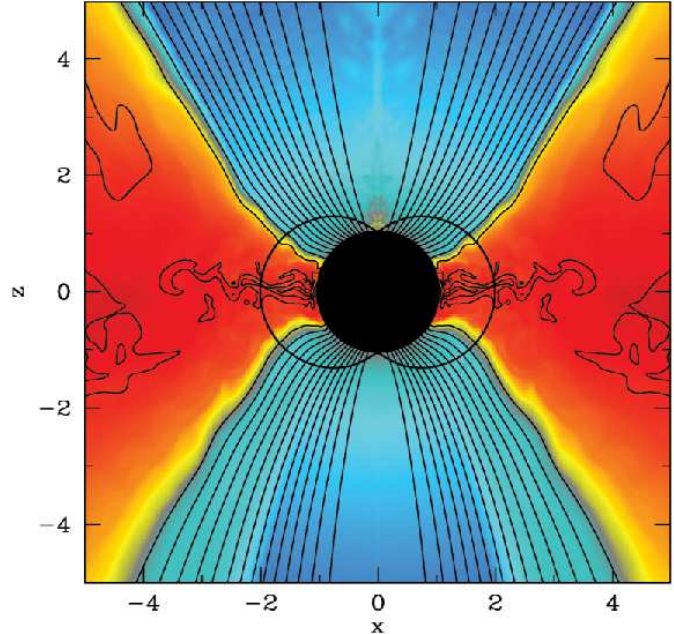}}
        \caption{Magnetic field lines and logarithm of rest-mass density (colored)
        quasi-steady  accretion disk simulations for an almost extreme Kerr black hole
        ($a/M=0.999$) [Fig. 3 of \cite{KK}].}
        \protect\label{mhd:Fig.8.2}
\end{figure}

We have already remarked earlier that even for very rapidly rotating
black holes the Blandford-Znajek process can drive magnetized jets.
Fig. \ref{mhd:Fig.8.2} from \cite{KK} shows clearly that there is no
``Meissner effect'' at work. This is due to the fact that within the
ergosphere the plasma unavoidably corotates with the black hole.

One of the main shortcomings of existing simulations is that the
radiation field is completely neglected. It is likely that radiative
and other high energy processes play a significant role in the flow
dynamics through radiation force on the outflowing plasma. Moreover,
at some point physical resistivity will also have to be included,
and the single-fluid approximation will not always hold everywhere.
Numerical calculations will be performed in 3D.

\appendix

\section{Numerical Schemes for GRMHD}\label{Sect:num}

Most GRMHD codes have adopted a \emph{conservative scheme}, which
means that the integrated evolutionary equations are of the general
form
\begin{equation}
\partial_t\underline{U}(\underline{P})+
\partial_i\underline{F}^i(\underline{P})=\underline{S}(\underline{P}).
\label{ap:1}
\end{equation}
Here, $\underline{U}$ are vector-valued ``conserved'' variables,
$\underline{P}$ is a vector of ``primitive'' variables (rest-mass
density, internal energy density, velocity components and magnetic
field components). The fluxes $\underline{F}^i,~\underline{U}$ and
the ``source vector'' $\underline{S}$ depend on $\underline{P}$.
Conservative numerical schemes advance $\underline{U}$, then
calculate $\underline{P}(\underline{U})$ once or twice per time
step.

While in praxis the map $\underline{P}\mapsto
\underline{U}(\underline{P})$ is analytically known, the inverse map
$\underline{U}\mapsto \underline{P}(\underline{U})$ is not available
in closed form and must be computed numerically. How this is
performed is at the heart of a conservative scheme, since the
operation must be accurate, fast and robust.

Below we describe these points in more detail, following \cite{NG}.
We use the notation and basic equations introduced in Sect. 5.

Equations (\ref{mhd:6.2}), (\ref{mhd:6.3}) and (\ref{mhd:6.7b}) can
readily be written in the form (\ref{ap:1}) in terms of the eight
conserved variables
\begin{equation}
\underline{U}:~~ D:=\gamma\rho_0,~~ Q_\mu:=-u_\nu T^\nu{}_\mu,~~
\vec{B}, \label{ap:2}
\end{equation}
where $\gamma=-u_\mu U^\mu=(1-v^2)^{-1/2}$. As eight primitive
variables we use
\begin{equation}
\underline{P}:~~\rho_0,~~ \varepsilon,~~ \vec{B},~~ v^i.
\label{ap:3}
\end{equation}
(The vector $\vec{B}$ is common to $\underline{U}$ and
$\underline{P}$.)

In \cite{NG} it is shown that $W:=\rho_0 h\gamma^2$, where
$h:=1+\varepsilon+p/\rho_0$, and $v^2$ can be determined from
$\underline{U}$ by solving two polynomial equations. After that one
can easily recover all primitive variables.

\section{Magneto-rotational Instability}

This important instability, whose crucial astrophysical implications
have been understood astonishingly late, shows up already in
linearized theory. For a detailed pedagogical discussion we refer to
\cite{BH}.

In a systematic treatment one linearizes the basic equations of MHD
for small fluctuations of a disk system, consisting of a central
point mass and a differentially rotating magnetized disk. An
analysis of the resulting rather complicated dispersion relation
leads to the following results. In the non-rotating limit
(homogeneous unperturbed situation) one finds the familiar
Alfv\'{e}n waves and two other MHD modes. The fast one is often
referred to as magnetosonic wave, and represents magnetic and
thermal pressure in concert. In the slow mode magnetic tension and
gas compression act in opposition. For weak fields it becomes
degenerate with the Alfv\'{e}n mode, while for strong fields it
becomes an ordinary sound wave, channeled along the field lines.

The effect of Kepler rotation on these three modes is very
interesting: At a critical rotation frequency the \emph{slow MHD
mode becomes unstable}.

There is a simple way to understand this far reaching instability.
Consider an axisymmetric gas disk in the presence of a vertical
magnetic field, that has no effect on the disk equilibrium. Assume
that a fluid element is displaced from its circular orbit by
$\mathbf{\xi} \propto e^{ikz}$ ($z=$ vertical direction). Using the
induction law and simple mechanics one finds the these
incompressible planar displacements satisfy the same equation as the
separation of two orbiting mass points, connected by a spring (with
a spring constant related to the Alfv\'{e}n velocity). It is quite
obvious that this system is unstable; the separation of the two mass
points rapidly increases. This is the essence of the weak-field
magneto-rotational instability.

The magneto-rotational instability plays a fundamental role in disk
accretion, because it leads to disk-turbulence and corresponding
stresses.


\begin{theacknowledgments}
I wish to thank the organizers of the Symposium on Gravitation and Cosmology, in particular
C. Laemmerzahl and A. Macias, for inviting me and their wonderful hospitality. I am grateful 
that J.C. McKinney and S.S. Komissarov allowed me to include two figures of their work in this article.
\end{theacknowledgments}



\bibliographystyle{aipproc}   


\begin{thebibliography}{9}
\bibitem{1} K.~S. Thorne, R. H. Price  \& D. A. MacDonald,
  \emph{Black Holes: The Membrane Paradigm}, Yale Univ. Press. (1986).

\bibitem{2} R. Durrer  \& N. Straumann, \emph{Helv. Phys. Acta.}
  \textbf{61}, 1027 (1988).

\bibitem{3} R. M. Wald, \emph{Phys. Rev.} \textbf{D 10}, 1680 (1974).

\bibitem{KK} S. S. Komissarov and J. C. McKinney,
  astro-ph/0702269.



\bibitem{6} K. S. Thorne \& D. A. MacDonald,
  \emph{Mon. Not. Roy. Astron. Soc.} \textbf{198}, 339 (1982).

\bibitem{7} N. Straumann, {\it General Relativity, With Applications to Astrophysics}, Texts and
Monographs in Physics, Springer Verlag, 2004.

\bibitem{8} N.~Straumann, The Membrane Model of Black Holes and
  Applications. In: F.W.~Hehl, C.~Kiefer and
  R.J.K.~Metzler (eds.), \textit{Black Holes: Theory and
  Observation}. Springer-Verlag 1998; astro-ph/9711276.

\bibitem{9} R. D. Blandford \& R. L. Znajek,
  \emph{Mon. Not. Roy. Astron. Soc.}
  \textbf{179}, 433 (1977).

\bibitem{10} J. C. McKinney and Ch. F. Gammie,
  \emph{Astrophys. J.}
  \textbf{611}, 977 (2004).



\bibitem{LWB} H. K. Lee, R. A. Wijers and G. Brown,
  \emph{Physics Reports}
  \textbf{325}, 83 (2000).

\bibitem{Ko} S. Koide,
  \emph{Phys. Rev.} \textbf{D 67}, 104010 (2003).

\bibitem{VH} J.-P. De Villiers \& J. F. Hawley,
  \emph{Astrophys. J.}
  \textbf{589}, 458 (2003).

\bibitem{Harm} Ch. F. Gammie, J. C. McKinney \& G. T\'{o}th,
  \emph{Astrophys. J.}
  \textbf{589}, 444 (2003).

\bibitem{Kom} S. S. Komissarov,
  \emph{MNRAS}
  \textbf{350}, 1431 (2004).

\bibitem{EH} C. R. Evans \& J. F. Hawley,
  \emph{Astrophys. J.}
  \textbf{332}, 659 (1988).

\bibitem{HK} J. F. Hawley \& J. H. Krolik,
  \emph{Astrophys. J.}
  \textbf{641}, 103 (2006) [astro-ph/0512227].

\bibitem{BH} S. A. Balbus \& J. F. Hawley,
  \emph{Rev. Mod. Phys.}
  \textbf{70}, 1 (1998).

\bibitem{KMS} Y. Kato, S. Mineshige \& K. Shibata,
  \emph{Astrophys. J.} \textbf{605}, 307 (2004).

\bibitem{NG} S. C. Noble, C. F. Gammi, J. C. McKinney, L. D. Del
Zanna
  \emph{Astrophys. J}
  \textbf{641}, 626 (2006) [astro-ph/0512420].

\bibitem{McK} J. C. McKinney,
  \emph{Mon.Not.R.Astron.Soc.} \textbf{368}, 1561 (2006) [astro-ph/0506369].

\bibitem{Kom2} S. S. Komissarov, M. V. Barkov, N. Vlahakis and A.
K\"{o}nigl,
  astro-ph/0703146.


\end{thebibliography}

\IfFileExists{\jobname.bbl}{}
 {\typeout{}
  \typeout{******************************************}
  \typeout{** Please run "bibtex \jobname" to optain}
  \typeout{** the bibliography and then re-run LaTeX}
  \typeout{** twice to fix the references!}
  \typeout{******************************************}
  \typeout{}
 }



\end{document}